\documentclass[aps,showpacs,PRL,twocolumn]{revtex4}
\usepackage{graphics}
\usepackage{graphicx}
\usepackage{amssymb}
\usepackage{wasysym}
\usepackage{pxfonts}
\usepackage{epsfig}
\usepackage{color}
\usepackage{ulem}
\newlength{\upit}\upit=0.1truein

\newcommand{\ltappr}{{{\lower4pt\hbox{$<$} } \atop \widetilde{ \ \ \ }}}
\newlength{\bxwidth}\bxwidth=1.5 truein

\newlength{\figwidth}
\newlength{\shift}
\newcommand \bea {\begin{eqnarray} }
\newcommand \eea {\end{eqnarray}}

\begin{document}

\title{Evidence for a Non-Fermi-Liquid Phase in Ge-Substituted YbRh$_2$Si$_2$}
\author{J.\ Custers$^{1}$,
P.\ Gegenwart$^{2,\S}$,
C.\ Geibel$^{2}$,
F.\ Steglich$^{2}$,
P.\ Coleman$^{3}$, and
S.\ Paschen$^{1}$}

\affiliation{$^1$Institute of Solid State Physics, Vienna University of
Technology, Wiedner Hauptstr.\ 8-10, 1040 Vienna, Austria}
\affiliation{$^2$Max Planck Institute for Chemical Physics of Solids, 01187
Dresden, Germany}
\affiliation{$^3$Center for Materials Theory, Rutgers University, Piscataway, NJ
08855, USA} 

\begin{abstract}
The canonical view of heavy fermion quantum criticality assumes a single quantum
critical point separating the paramagnet from the antiferromagnet. However,
recent experiments on Yb-based heavy fermion compounds suggest the presence of
non-Fermi liquid behavior over a finite zero-temperature region. Using detailed
susceptibility and transport measurements we show that the classic quantum
critical system, Ge-substituted YbRh$_2$Si$_2$, also displays such behavior. We
advance arguments that this is not due to a disorder-smeared quantum critical
point, but represents a new class of metallic phase. 
\end{abstract}

\maketitle
\newcommand{\YRS}{YbRh$_2$Si$_2$}
\newcommand{\YIS}{YbIr$_2$Si$_2$}
\newcommand{\gYRS}{YbRh$_{2}$(Si$_{1-x}$Ge$_{x}$)$_2$}
\newcommand{\YRIS}{Yb(Rh$_{0.94}$Ir$_{0.06}$)$_2$Si$_{2}$}
\newcommand{\YRSG}{YbRh$_{2}$(Si$_{0.95}$Ge$_{0.05}$)$_2$}
\newcommand{\YAG}{YbAgGe}
\newcommand{\YBAL}{YbAlB$_4$}

Quantum criticality in heavy fermion (HF) compounds has been a topic of great
interest for more than a decade~\cite{Loe07.1s}. In the vicinity of a quantum
critical point (QCP), HF materials display qualitative departures from the
standard Landau Fermi liquid (LFL) behavior of conventional metals over a wide
temperature ($T$) range. Our failure to understand these phenomena constitutes a
major unsolved problem in physics.

A key element in the debate about quantum criticality is whether the quantum
critical physics of HF materials can be understood using a space-time
generalization of classical criticality, often called ``Hertz-Millis''
theory~\cite{Her76.1,Mil93.1}, or whether a new framework, evoking the critical
breakdown of Kondo screening at the quantum critical
point~\cite{Col01.1s,Si01.1s,Sen04.1,Sen04.2s,Pep07.1} is required. Recently,
however, a new issue has arisen. Experiments on Yb-based HF systems, including
\YRIS~\cite{Fri09.1s}, YbAgGe~\cite{Bud04.1}, and
$\beta$-YbAlB$_4$~\cite{Nak08.1s} have observed the presence of non-Fermi liquid
(NFL) behavior over a finite zero-$T$ region of the magnetic field ($B$)- or
pressure ($p$)-tuned phase diagram, rather than at a single QCP. These
observations raise the possibility that our underlying scenario for HF quantum
criticality may need to be changed.

In this letter we re-investigate the low-$T$ magnetotransport and
magnetic susceptibility properties of \YRSG~\cite{Cus03.1s} to examine the
earlier assumption of a single QCP. Together with
CeCu$_{5.9}$Au$_{0.1}$~\cite{Sch00.1s} this material has played a major role in
establishing the central properties of HF quantum criticality.

The $T$-$B$ phase diagram of the mother compound \YRS\ represents,
with a well defined fan of NFL behavior in a LFL background~\cite{Cus03.1s}, a
particularly clear example of a quantum critical {\it point}. When a nominal
concentration of 5~at\% Si is substituted by Ge~\cite{YRSGfootnote1s}, this QCP
was observed to move, for $B \perp c$, from 0.06~T further down
towards 0.025~T~\cite{Cus03.1s}. Thus, a material situated, at $B=0$, extremely
close to its QCP appeared to be found. As will be shown below the NFL behavior
of \YRSG\ no longer emerges from a single point but occupies a finite segment of
the $B$ axis at $T = 0$. In addition the Kondo breakdown scale
$T^{\ast}$~\cite{Pas04.1s,Geg07.1s} is now detached from the zero $T$
magnetic phase transition. This observation is of great interest in the context
of new results for Co- and Ir-substituted \YRS~\cite{Fri09.1s} as well as on
CeIn$_3$ under high $B$~\cite{Seb09.1s}, which also indicate a
separation of $T^{\ast}$ and $T_{\mathrm{N}}$ at zero $T$.

\begin{figure}[t]
\centerline{\includegraphics[clip,width=0.64\columnwidth,angle=-90]{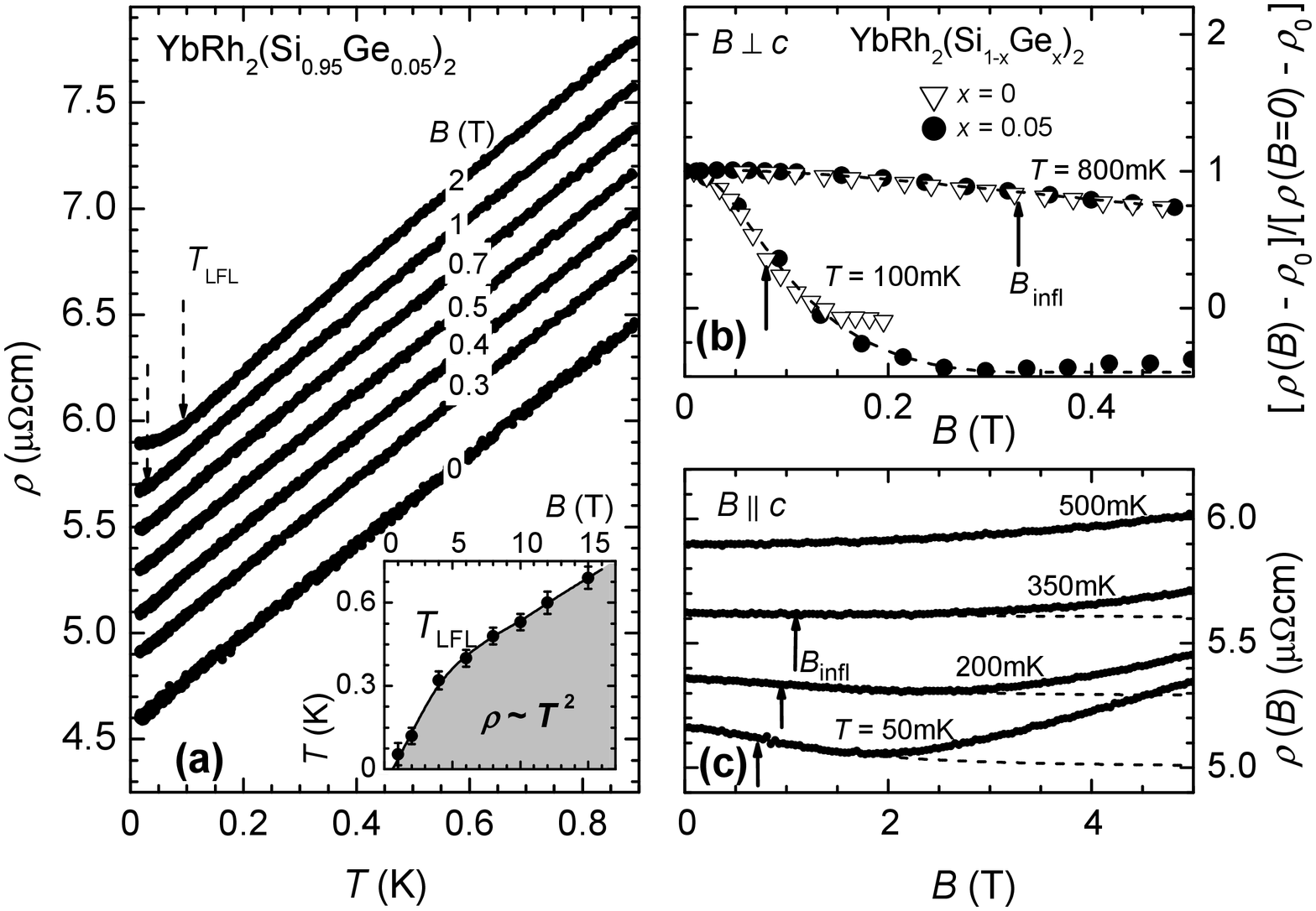}}
\caption{\label{rho} Electrical resistivity of \YRSG. (a) Low-$T$
resistivity for various $B \parallel c$. For clarity curves at $B > 0$ are
shifted by 0.2~$\mu\Omega$cm with respect to each other. Below
$T_{\mathrm{LFL}}$, LFL behavior is recovered. Inset: Shaded area depicts region
of LFL behavior determined from $\rho(T)$. The $T_{\mathrm{LFL}}$ line is a
polynomial fit and serves as guide-to-the-eye. (b) Scaled isothermal
magnetoresistance for $B \perp c$ ($\rho_0 = 4.51 \mu\Omega$cm). Data for pure
\YRS\ (open symbols, $\rho_0 = 1.81 \mu\Omega$cm) are shown for comparison. (c)
Isothermal magnetoresistance for $B \parallel c$. The arrows denoted by
$B_{\mathrm{infl}}$ mark the inflection points~\cite{Geg07.1s} of the fits to
the data (full and dashed lines in (a) and (b), respectively) using the
crossover function introduced in Ref.~\onlinecite{Pas04.1s}.}
\end{figure}

In Fig.\,\ref{rho}(a) we show the $T$ dependence of the electrical
resistivity, $\rho(T)$, of \YRSG\ for $B \parallel c$ up to 2~T. A
striking linearity of $\Delta\rho(T)$ between at least 20~mK and 900~mK develops
over the $B$ range from zero to above 0.7~T. To our knowledge, such
robust NFL behavior in a finite $B$ range (factor of 45 in $T$, factor
of 2.3 in $B$ considering the expected crossover to a quadratic dependence at
even lower $T$ in the AFM state at $B < B_{\mathrm{c1}} = 0.3$~T, see
Fig.\,\ref{phasedia}) has not previously been observed in any HF
compound~\cite{Bud04.1,Fri09.1s,Nak08.1s}. In analogy with \YRS~\cite{Geg07.1s},
the resistivity $\rho(B)$ isotherms have been examined. Clear crossover behavior
is seen for $B \perp c$ and $B \parallel c$ which is characterized by inflection
points~\cite{Geg07.1s} denoted as $B_{\mathrm{infl}}$ in Fig.\,\ref{rho}(b) and
Fig.\,\ref{rho}(c), respectively. It is clear from these figures that
$B_{\mathrm{infl}}$ increases with increasing $T$. Like Co- and
Ir-substituted \YRS~\cite{Fri09.1s}, the crossover behavior for the
Ge-substituted compound investigated here is found to be almost identical with
the one of pure \YRS\ (Fig.\,\ref{rho}(b)).

\begin{figure}[t]
\centerline{\includegraphics[clip,width=0.82\columnwidth]{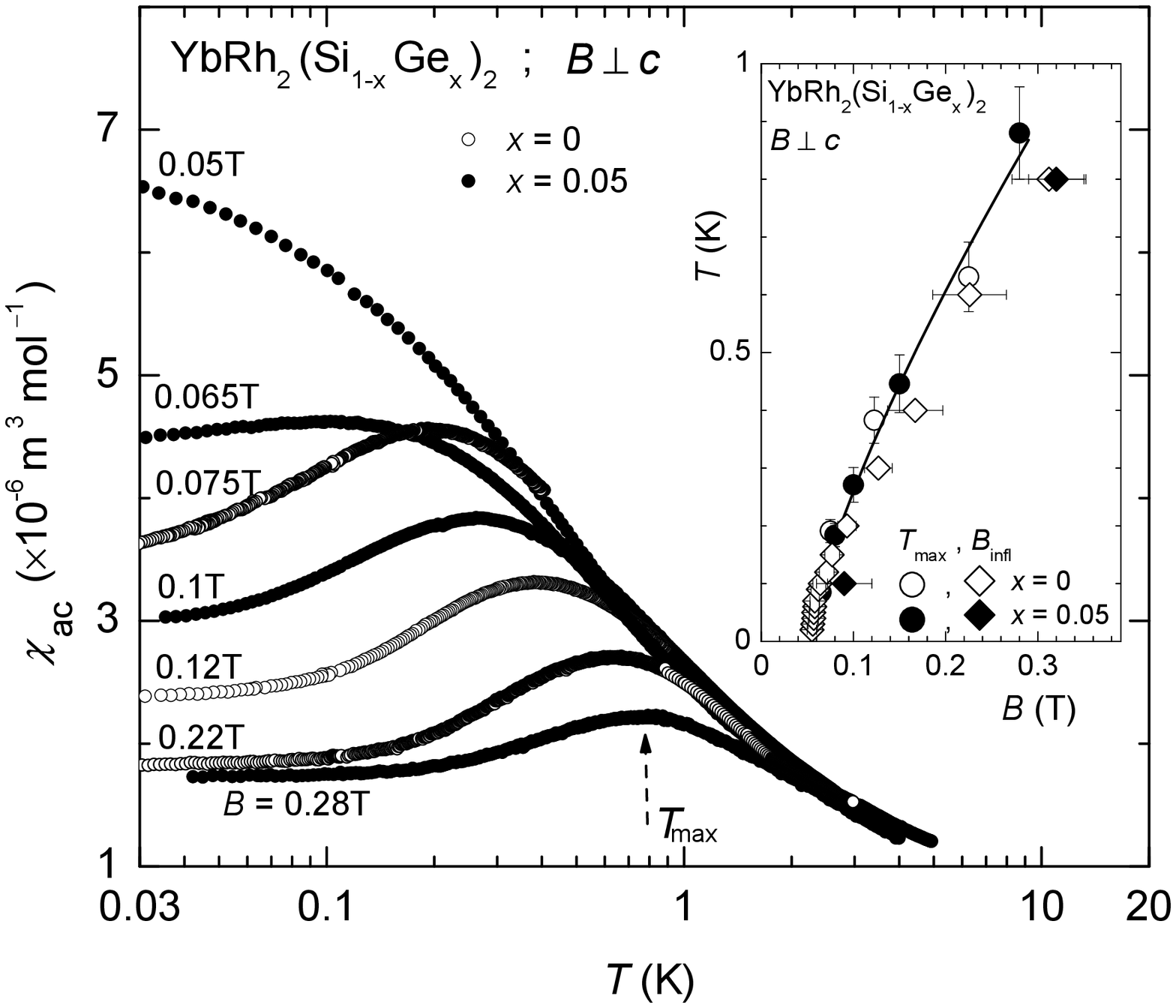}}
\caption{\label{chi} Ac susceptibility of \gYRS\ for $x =
0.05$ (full symbols) and, for comparison, for $x = 0$ (open symbols). Inset:
Positions of the maxima in iso-$B$ $\chi_{\mathrm{ac}}(T)$ curves (circles)
and of the inflection points~\cite{Geg07.1s} $B_{\mathrm{infl}}$ of $\rho(B)$ isotherms
(diamonds) of both samples. The power law $(B-B_{\mathrm{c}})^{0.75}$ (solid line)
with $B_{\mathrm{c}} = 0.06$~T is a good description of {\it all} data points.}
\end{figure}

\begin{figure}[t]
\centerline{\includegraphics[clip,width=0.68\columnwidth,angle=-90]{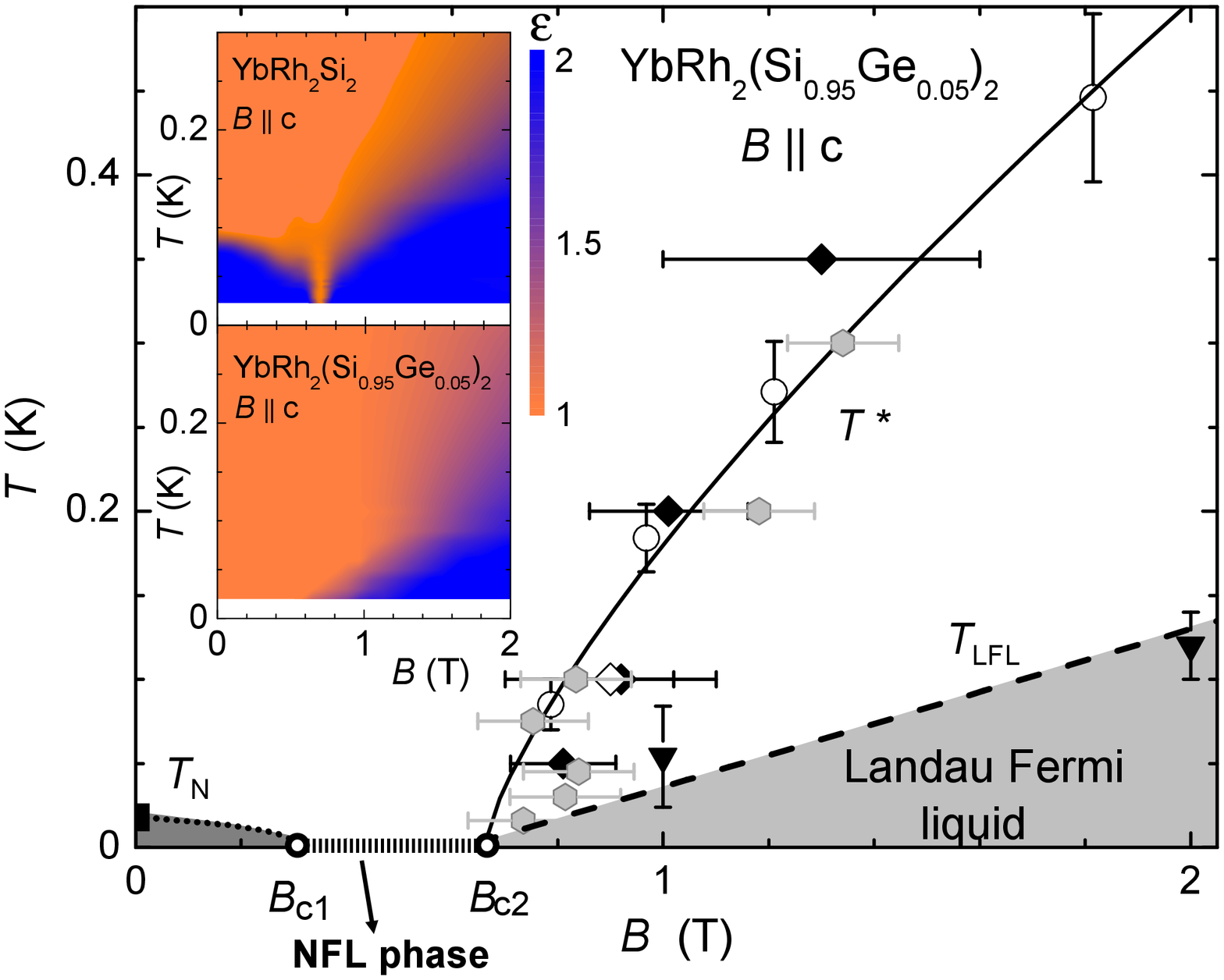}}
\caption{\label{phasedia} (Color online) Phase diagram of \YRSG\
for $B \parallel c$. Symbols represent $B_{\mathrm{infl}}$ ($\Diamondblack$) and
the upper boundary of LFL behavior ($\blacktriangledown$). The dashed
$T_{\mathrm{LFL}}$ line is the polynomial fit shown in the inset of
Fig.\,\ref{rho}(a). Data points from measurements with $B \perp c$ are included
by multiplying $B$ with the factor 11: $\Diamond$ symbolizes
$B_{\mathrm{infl}}$, $\Circle$ displays $T_{\mathrm{max}}$ from
$\chi_{\mathrm{ac}}(T)$. The solid $T^{\ast}$ line is taken from the inset of
Fig.\,\ref{chi}.  Hexagons represent $T^{\ast}$~\cite{Geg07.1s}
(or $T_{\mathrm{Hall}}$~\cite{Pas04.1s}) of \YRS. $\blacksquare$ marks
$T_{\mathrm N}$ observed by specific heat. The dotted $T_{\mathrm N}$ line
indicates the typical evolution of $T_{\mathrm N}$ for \YRS, $T_{\mathrm N}(B) =
T_{\mathrm N}(0)(1-B/B_{\mathrm c})^{0.36}$~\cite{Fri09.1s}, using the
respective parameters for \YRSG\ ($T_{\mathrm N}(0) = 18$~mK, $B_{\mathrm c} =
11 \times B_{\mathrm c}^{\perp c} \approx 0.3$~T)~\cite{Cus03.1s}. The hatched
area 0.3~T~$\le B \le $~0.66~T marks the zero $T$ NFL phase
characterized by $\Delta\rho\sim T$. The inset compares the evolution of the
resistivity exponent $\varepsilon$, derived from the dependence $(\rho - \rho_0)
\sim T^\varepsilon$ (see also Ref.~\onlinecite{Cus03.1s}), for \YRS\ (top) and
\YRSG\ (bottom) in the same $B$ and $T$ range.}
\end{figure}

This is further supported by another measure of the crossover scale $T^{\ast}$,
the position $T_{\mathrm{max}}$ of maxima in iso-$B$ $\chi(T)$
curves~\cite{Geg07.1s}, cf.\, Fig.\,\ref{chi}. Like $\rho(B)$, also the $\chi(T)$
data show that, while $T_{\mathrm{N}}$ is strongly suppressed upon substituting
\YRS\ with Ge, $T^{\ast}$ does not move (Fig.\,\ref{chi}, inset).

Figure~\ref{phasedia} summarizes all characteristic features of \YRSG\ in a
$T$-$B$ phase diagram. As indicated by the shaded area a
finite range of NFL behavior at zero $T$ appears between the critical
fields $B_{\mathrm{c1}}$ and $B_{\mathrm{c2}}$ for the suppression of
$T_{\mathrm{N}}$ and $T^{\ast}$.

In pure \YRS, the in-$T$ linear resistivity extends to the lowest accessible
$T$ (20~mK) at a single critical $B$, yet in \YRSG\ this canonical
behavior is violated, and instead, in-$T$ linear resistivity extends to the
lowest $T$ over a substantial $B$ range. In isolation, this behavior
might be dismissed as an anomaly. However, similar behavior has recently been
observed also in other Yb-based HF
compounds~\cite{Fri09.1s,Bud04.1,Nak08.1s}.

Conservatively, we might attribute these observations to disorder. In the
Hertz-Millis theory, the in-$T$ linear resistivity of HF systems is itself
attributed to disorder~\cite{Ros99.1,Mor95.1}. Furthermore, disorder is expected
to smear a well-defined QCP into a region~\cite{Hoy08.1}.

However, various aspects speak against this conservative view point. Firstly, it
is unlikely that the smearing of a QCP will be ``asymmetric''. The position of
the $T^\ast$-line in \gYRS\ and hence of the entrance into the LFL phase is not
affected by going from $x=0$ to $x=0.05$ (see
Refs.\,~\onlinecite{Pas04.1s,Geg07.1s} for the phase diagram of \YRS); the NFL
region in \YRSG\ thus spreads only to the left of of $T^\ast$. Secondly, the NFL
power law dependencies are identical for \YRSG\ and \YRS~\cite{Cus03.1s}. Thus,
either both systems are disorder dominated or none. And finally, values for the
normalized linear rise of resistivity $\Delta\rho/\rho_0$ are, with $\approx 4$
for \YRSG~\cite{Tro02.1s}, $\approx 5$ for early \YRS\ samples~\cite{Tro00.2s},
and $\approx 20$ for the new generation of ultrapure \YRS\
(where $\rho = \rho_0 + A T^{\alpha}$ with $\alpha = 1 \pm 0.2$ holds up to
20~K)~\cite{Wes09.1}, all beyond the maximum value of unity expected within the
Hertz-Millis type scenario for disordered systems~\cite{Ros99.1}.
$\Delta\rho/\rho_0$ values more compatible with this scenario are observed for
CeCu$_{5.9}$Au$_{0.1}$ ($\Delta\rho/\rho_0\approx 0.5$)~\cite{Loe94.1s} and
YbAgGe ($\Delta\rho/\rho_0\approx 1$)~\cite{Bud04.1}, values much larger than
unity for CeCoIn$_{5}$ ($\Delta\rho/\rho_0\approx 100$ for $I \perp
c$)~\cite{Tan01.1s}. Of course, the significance of $\Delta\rho/\rho_0$ in
estimating the role of disorder is questionable in systems as \YRS\ and \YRSG\
where the Hertz-Millis theory fails~\cite{Cus03.1s,Pas04.1s,Geg07.1s}.

A natural interpretation of our data, then, is that the $B$-region with in-$T$
linear resistivity corresponds to a well-defined metallic phase with
unconventional transport properties. The possibility of such ``strange metal''
phases that lie beyond a conventional LFL description has been
discussed in a variety of theoretical
contexts~\cite{Sen08.1,Pep07.1,And08.1,Nev09.1}. One line of reasoning argues
that the transition into the strange metal phase involves a partial Mott
localization of the $f$ quasiparticle degrees of freedom. Such a view is 
consistent with the $T^{\ast}$ line that connects to the upper $B$ edge of the
strange metal phase. Anderson~\cite{And08.1} has recently proposed the
possibility of a ``hidden Fermi liquid'', in which well-defined quasiparticles
are present, but can not be created singly by the addition of external
electrons.

\begin{figure}[t]
\centerline{\includegraphics[clip,width=0.65\columnwidth,angle=-90]{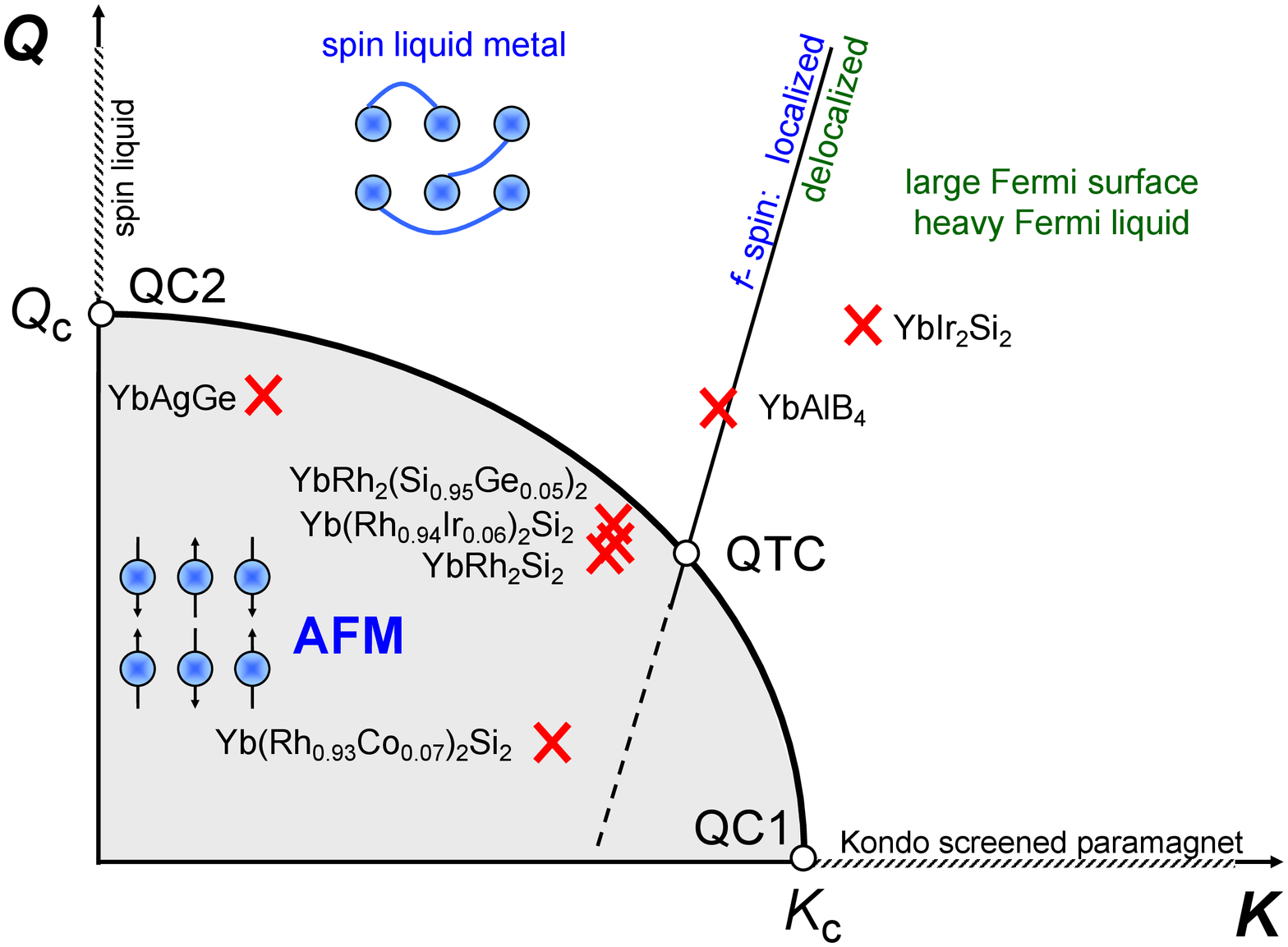}}
\caption{\label{theory} (Color online) Generic phase diagram displaying the
combined effects of Kondo coupling ($K$) and magnetic frustration, or quantum
zero point motion ($Q$). For the location of compounds in the phase
diagram (red crosses), see text.}
\end{figure}

The standard model of HF physics is based on Doniach's ``Kondo
lattice hypothesis'', according to which deviations from the HF
magnetic QCP are tuned by a single parameter
$K=T_{\mathrm{K}}/J_{\mathrm{H}}$, the ratio of the single-ion Kondo temperature
$T_{\mathrm{K}}$ to the nearest neighbor RKKY interaction $J_{\mathrm{H}}$. As
$K$ is increased, a quantum phase transition QC1 takes place at some value
$K_{c}$ (see abscissa of Fig.\,\ref{theory}, $Q=0$).

The apparent emergence of a strange metal phase in certain HF
compounds leads us to propose a two parameter extension to the Doniach phase
diagram which considers the interplay of the Kondo effect ($K$) with the quantum
zero point motion of the local moments ($Q$). Related ideas have been previous
considered by various authors~\cite{Bur02.1,Si06.1,Leb07.1,Voj08.1,Ong09.1}. To
motivate this idea, we first consider a ``drained Kondo lattice'' ($K=0$), with
only local moments coupled together on a lattice by a short-range
antiferromagnetic (AFM) Heisenberg interaction. In isolation this lattice would
develop an AFM ground state. However, by adding a frustrated
second-neighbor coupling between the spins or, more abstractly, by reducing the
size of the magnetic moment we can increase the strength of the quantum
zero-point spin fluctuations. At some critical value $Q_{c}$, there is then a
quantum phase transition QC2 where long-range magnetic order melts under the
influence of zero-point spin fluctuations~\cite{Cha88.1} to form a spin liquid
(see ordinate of Fig.\,\ref{theory}).

For the general case $K\neq 0$, $Q\neq 0$ we may link QC1 and QC2 via a single
phase boundary (Fig.\,\ref{theory}). However, this line is not the only feature
in the phase diagram. When we turn on a small Kondo coupling between the
conduction electrons and the spin liquid, the Kondo effect will not turn on
instantly, since the spin liquid has a characteristic energy scale which will
cut-off the Kondo logarithms. So, for small Kondo coupling, the spin liquid will
co-exist with a small Fermi surface metal to form a ``spin liquid metal''. The
concept of a Kondo-stabilized spin liquid metal was first proposed in
Ref.\,\onlinecite{Col89.1}, but has more recently been discussed in connection
with frustrated Kondo lattices~\cite{Nak06.1s}, and as topologically distinct
LFL phase of the Kondo lattice~\cite{Voj99.1,Sen04.1}. Since the volume of the
Fermi surface in the spin liquid metal is an invariant, the small and large
Fermi surface states are topologically distinct phases, separated by at least
one quantum phase transition. In the simplest scenario, a single quantum phase
transition from a small to a large Fermi surface must take place at some
critical Kondo coupling $K_{c} (Q)$. In this way, the generalized magnetic-Kondo
phase diagram must contain two {\it independent} quantum-critical lines - one
where long range order develops, and another where the volume of the Fermi
surface jumps. In general, these two lines will cross at a quantum
tetra-critical point (QTC), where the $f$ electrons localize at the same time as
magnetic order develops. Tuning the parameters of the material in the vicinity
of this QTC will, in this scenario, cause the two transitions to separate: for
transitions that take place at $Q$ values above the QTC, the AFM QCP will have
the character of a localized magnetic QCP, whereas for $Q$ values that lie below
the QTC, the QCP will have the character of an itinerant (spin density wave)
transition. Another interesting possibility which has been discussed is that,
instead of a QTC, a line exists in the $Q-K$ phase diagram on which the
selective Mott transition and the AFM quantum critical point
coincide~\cite{Si06.1}.

In real HF materials, the precise relationship of $p$ and $B$ tuning the $Q$ and
$K$ axes in our diagram will depend on microscopic details. For each material,
$p$ and $B$ will define two generally non-orthogonal, but independent directions
in the $Q-K$ plane. Most of the existing data on Yb quantum critical systems
have been interpreted assuming that $B$ tuning can be identified with the $K$
axis. The independent effect of doping or $p$ must therefore at least have a
finite component along the $Q$ direction. With these assumptions, we can
tentatively locate various Yb-based HF compounds in the $Q-K$ plane.

According to its $T-B$ phase diagram (Fig.\,\ref{phasedia}) \YRSG\ is situated
in the AFM region, to the left of the localized-to-delocalized line. The low
value of $T_{\mathrm{N}}$ (and the correspondingly low value of
$B_{\mathrm{c1}}$) suggests that it is close to the AFM line. \YRS\ lies just to
the left of the QTC in the phase diagram~\cite{Pas04.1s}, while \YRIS\ lies, due
to its smaller $T_{\mathrm{N}}$, at a higher values of $Q$~\cite{Fri09.1s}, so
that $B$ tuning into the paramagnet takes place via an intermediate spin
liquid metal. By contrast, Co-substituted \YRS\ with a larger $T_{\mathrm{N}}$
lies at a lower value of $Q$, so that $B$ tuning induces a delocalization of
the $f$ quasiparticles before the loss of magnetic order. The fact that
$B_{\mathrm{c2}}$ is essentially the same for all these compounds is accounted
for by positioning them around a line parallel to the localized-to-delocalized
line. Their sequence along this line and their relative distances reflect the
decrease of the chemical $p$ (increase of the unit cell volume) towards the
top~\cite{Cus03.1s,Mac08.1s,Fri09.1s}.

The ground state of pure \YIS\ is a LFL. At the lowest $T$ no signature
of $T^\ast$ was found in $B$-dependent Hall effect
measurements~\cite{Kri08.1s}. In addition $p$ experiments revealed that the
paramagnetic ground state is stable up to $p \approx 8$~GPa~\cite{Yua06.1s}
leading us to place \YIS\ tentatively to the right of the $f$-spin localization
line. We can also incorporate the parallel observations on \YAG\ and \YBAL\ into
this framework. \YBAL\ is seen to enter a LFL phase almost immediately upon
application of $B$~\cite{Nak08.1s}, suggesting that this system lies just at
the edge of the spin liquid metal phase. \YAG\ requires a substantial $B$ for
the destruction of the antiferromagnetism, but beyond the AFM QCP, it is seen to
pass through a finite $B$ range with linear resistivity (NFL
region)~\cite{Bud04.1}. High-$B$ effects like the putative Lifshitz transition
in \YRS~\cite{Rou08.1s} at about 10~T fall outside the validity range of the
phase diagram.

A central question raised by this discussion concerns the excitations of the
proposed spin liquid metal. In particular, is the transport of charge, spin
and/or heat carried by coherent fermions, or is a more radical description
required?  More detailed transport and spectroscopic measurements on Yb-based
HF compounds with putative NFL
phases~\cite{Fri09.1s,Bud04.1,Nak08.1s} are clearly needed to help characterize
the excitations in the $B$ regime of linear resistance. It is equally
important to experimentally quantify the theoretical parameters $K$ and $Q$.
This might be achieved with neutron scattering experiments in which the momentum
independent quasielastic linewidth is a measure of the strength of the Kondo
effect ($K$) while the inelastic response at specific magnetic $q$ vectors is a
measure of the quantum zero point motion associated with antiferromagnetism
($Q$)~\cite{Mon03.1s}.

In conclusion, we have presented a set of transport and susceptibility
measurements which strongly suggest the presence of a new metallic phase in
magnetic field-tuned \YRSG, nested between the antiferromagnetic and LFL ground states of
this material. We have proposed a two-dimensional generalization of the Doniach
phase diagram as a framework for interpreting this, and other recently observed
cases of strange metal phases of similar character. Future experiments, based on
pressure tuning could provide an important means of veryfing these new ideas.

We gratefully acknowledge discussions with A.\ Nevidomskyy, S.\ Nakatsuji, and
A.\ Millis. This work was supported by the ERC Advanced Grant n$^{\circ}$~227378
(SP) and by NSF DMR 0907179 (PC).

\footnotesize{\noindent$^{\S}$Present address: I.\ Physik.\ Institut,
Georg-August-Universit\"at G\"ottingen, 37077 G\"ottingen, Germany.}
\vspace{-0.4cm}


\begin{thebibliography}{10}

\bibitem{Loe07.1s}
{H.\ {v. L\"ohneysen} et al.}, {Rev.\ Mod.\ Phys.} {\bf 79},  1015  (2007).

\bibitem{Her76.1}
J. Hertz, {Phys.\ Rev.\ B} {\bf 14},  1165  (1976).

\bibitem{Mil93.1}
A.~J. Millis, {Phys.\ Rev.\ B} {\bf 48},  7183  (1993).

\bibitem{Col01.1s}
{P.\ Coleman et al.}, {J.\ Phys.: Condens.\ Matter} {\bf 13},  R723  (2001).

\bibitem{Si01.1s}
{Q.\ Si et al.}, Nature {\bf 413},  804  (2001).

\bibitem{Sen04.1}
T. Senthil, M. Vojta, and S. Sachdev, {Phys.\ Rev.\ B} {\bf {69}},  035111
  ({2004}).

\bibitem{Sen04.2s}
{T.\ Senthil et at.}, {Science} {\bf {303}},  1490  ({2004}).

\bibitem{Pep07.1}
C. P{\'e}pin, {Phys.\ Rev.\ Lett.} {\bf 98},  206401  (2007).

\bibitem{Fri09.1s}
{S.\ Friedemann et al.}, {Nature Phys.} {\bf {5}},  {465}  ({2009}).

\bibitem{Bud04.1}
S.~L. Bud\char39{}ko, E. Morosan, and P.~C. Canfield, {Phys.\ Rev.\ B} {\bf
  69},  014415  (2004).

\bibitem{Nak08.1s}
{S.\ Nakatsuji et al.}, {Nature Phys.} {\bf 4},  603  (2008).

\bibitem{Cus03.1s}
{J.\ Custers et al.}, Nature {\bf 424},  524  (2003).

\bibitem{Sch00.1s}
{A. Schr\"{o}der et al.}, {Nature} {\bf 407},  351  (2000).

\bibitem{YRSGfootnote1s}
The actual Ge-content of our single crystal is only about 2~at\% (see caption
  of Fig.\ 1 in Ref.\,\onlinecite{Cus03.1s}).

\bibitem{Pas04.1s}
{S.\ Paschen et al.}, {Nature} {\bf 432},  881  (2004).

\bibitem{Geg07.1s}
{P.\ Gegenwart et al.}, {Science} {\bf 315},  969  (2007).

\bibitem{Seb09.1s}
{S.\,E.\ Sebastian et al.}, {PNAS} {\bf {106}},  {7741}  (2009).

\bibitem{Ros99.1}
A. Rosch, {Phys.\ Rev.\ Lett.} {\bf 82},  4280  (1999).

\bibitem{Mor95.1}
T. Moriya and T. Takimoto, {J.\ Phys.\ Soc.\ Jpn.} {\bf 64},  960  (1995).

\bibitem{Hoy08.1}
J.~A. Hoyos and T. Vojta, Phys.\ Rev.\ Lett. {\bf 100},  240601  (2008).

\bibitem{Tro02.1s}
{O.\ Trovarelli et al.}, Physica B {\bf {312-313}},  {401}  ({2002}).

\bibitem{Tro00.2s}
{O.\ Trovarelli et al.}, {Phys.\ Rev.\ Lett.} {\bf 85},  626  (2000).

\bibitem{Wes09.1}
{T.~Westerkamp}, dissertation, {TU Dresden, Germany}, {2009}.

\bibitem{Loe94.1s}
{H.\ {v.\ L\"ohneysen} et al.}, {Phys.\ Rev.\ Lett.} {\bf 72},  3262  (1994).

\bibitem{Tan01.1s}
{X.\ Tang et al.}, {J.\ Mater.\ Res.} {\bf 16},  837  (2001).

\bibitem{Sen08.1}
T. Senthil, {Phys.\ Rev.\ B} {\bf {78}},  035103  ({2008}).

\bibitem{And08.1}
{P.\,W.\ Anderson}, {Phys.\ Rev.\ B} {\bf {78}},  {174505}  ({2008}).

\bibitem{Nev09.1}
A.~H. Nevidomskyy and P. Coleman, {Phys.\ Rev.\ Lett.} {\bf 102},  077202
  (2009).

\bibitem{Bur02.1}
S. Burdin, D.~R. Grempel, and A. Georges, Phys.\ Rev.\ B {\bf {66}},  {045111}
  ({2002}).

\bibitem{Si06.1}
Q. Si, {Physica B} {\bf {378-380}},  {23}  ({2006}).

\bibitem{Leb07.1}
E. Lebanon and P. Coleman, {Phys.\ Rev.\ B} {\bf {76}},  {085117}  (2007).

\bibitem{Voj08.1}
M. Vojta, {Phys.\ Rev.\ B} {\bf {78}},  {144508}  (2008).

\bibitem{Ong09.1}
T.~T. Ong and B.~A. Jones, {Phys.\ Rev.\ Lett.} {\bf {103}},  {066405}  (2009).

\bibitem{Cha88.1}
{See, for example, P.\ Chandra and B.\ Doucot}, {Phys.\ Rev.\ B} {\bf {38}},
  {9335}  ({1988}).

\bibitem{Col89.1}
P. Coleman and N. Andrei, {J.\ Phys.: Condens.\ Matter} {\bf {1}},  {4057}
  ({1989}).

\bibitem{Nak06.1s}
{S.\ Nakatsuji et al.}, {Phys.\ Rev.\ Lett.} {\bf {96}},  {087204}  ({2006}).

\bibitem{Voj99.1}
M. Vojta and S. Sachdev, {Phys.\ Rev.\ Lett.} {\bf {83}},  {3916}  (1999).

\bibitem{Mac08.1s}
{M.\,E.\ Macovei et al.}, {J.\ Phys.: Condens.\ Matter} {\bf 20},  505205
  (2008).

\bibitem{Kri08.1s}
{M.\ Kriegisch et al.}, Physica B {\bf 403},  1295  (2008).

\bibitem{Yua06.1s}
{H.\,Q.\ Yuan et al.}, {Phys.\ Rev.\ B} {\bf 74},  212403  (2006).

\bibitem{Rou08.1s}
{P.\,M.\,C.\ Rourke et al.}, {Phys.\ Rev.\ Lett.} {\bf 101},  237205  (2008).

\bibitem{Mon03.1s}
{W.\ Montfrooij et al.}, {Phys.\ Rev.\ Lett.} {\bf 91},  087202  (2003).

\end{thebibliography}

\end{document}